\documentclass[twocolumn]{aastex701}
\usepackage{amsmath}
\usepackage{booktabs}
\usepackage{array}
\usepackage{hyperref}

\newcommand{\nz}{\nu_0}
\newcommand{\gh}{\hat{g}}

\newcommand{\Pow}{\mathcal{P}}

\begin{document}

\title{Bicoherence States from QPO--Accretion Flow Coupling in Black Hole X-ray Binaries}

\author[orcid=0000-0003-3982-1466,sname='Seth Larner']{Seth Rowan Larner}
\affiliation{Department of Physics, Washington University in St.~Louis, St.~Louis, MO 63130, USA}
\email[show]{larner@wustl.edu}

\begin{abstract}

We calculate the bispectrum of a black hole X-ray binary light curve model where lognormal fluctuations associated with the inner flow are multiplicatively modulated by a quasi-periodic oscillation (QPO) and its harmonic. 
The bispectrum is a third-order Fourier statistic which measures the strength of nonlinear relationships between frequencies in a timeseries. When applied to black hole X-ray binaries, the bispectrum reveals nonlinear coupling between QPOs and the broadband noise. These couplings are sorted into a small set of states and X-ray binaries are observed to transition between these states as they undergo spectral transitions. We show that the model we consider is able to reproduce all of the canonical bispectral states analytically. We consider other sources of incoherent variability which dilute this model light curve and find that changes in this dilution, together with changes in the QPO amplitude, drive the transitions between the bispectral states. We also show that the larger QPO amplitudes of high-inclination sources favor the cross state and the weaker QPOs of low-inclination sources favor the hypotenuse state, which can explain the empirically observed correlation. Additionally, any dilution by equatorial winds, which are seen mainly at high inclination, would push high-inclination sources further toward the cross state.

\end{abstract}


\section{Introduction}
The X-ray light curves of black hole X-ray binaries (BHXRB) are highly dynamic on short timescales, frequently exhibiting both strong broadband noise and quasi-periodic oscillations (QPOs). QPOs manifest as sharp peaks in the X-ray power spectrum. They are typically categorized into types A, B, and C according to their strength, central frequency $(\nz)$, quality factor $(Q = \nz/\mathrm{FWHM})$, and the accompanying variability and spectral properties of the source \citep[types introduced by][and formalized by \citealt{Casella2005}]{Wijnands1999, Remillard2002}.\footnote{We focus in this work only on low-frequency QPOs as opposed to high-frequency QPOs with $\nz>100$~Hz in BHXRB \citep[see][]{Belloni2012}.}  

Type A QPOs appear as weak broad peaks in some high soft state power spectra. 
Type B QPOs are stronger and narrower than type A QPOs. They typically exhibit central frequencies $\nz \sim 5$--$6$~Hz and have quality factors $Q\gtrsim 6$ \citep{Casella2005}. Type B QPOs appear in the soft intermediate state, accompanied by weak broadband noise. Type B QPOs are often accompanied by peaks at the second harmonic of the main feature and sometimes the subharmonic \citep{Casella2005}. Finally, type C QPOs are the strongest and highest-$Q$ low-frequency QPOs. They appear at a broad range of central frequencies ($\nz \sim 0.1$--$10$~Hz) in the hard and hard intermediate states. They are accompanied by strong broadband noise, with total RMS amplitudes in this state reaching $30\%$. These QPOs are also frequently accompanied by harmonics and subharmonics.

The physical mechanism(s) which produce QPOs are still under debate. Models can generally be sorted into those which invoke an intrinsic change to the emission mechanisms in the source \citep[e.g., feedback between the disk and Comptonized coronal photons,][]{Bellavita2022} and those which describe a geometric change to the system \citep[e.g., Lense-Thirring precession of an inner accretion flow,][]{Stella1998, Ingram2009, Marcel2026}. 

Fundamentally, many models of QPO origin can reproduce the second-order (power spectral) behavior seen in type A, B, and C QPOs. This motivates the need to look beyond second order and examine the presence and effects of nonlinear variability in BHXRB\@. It is known that the light curves of BHXRB are nonlinear and described by a lognormal flux distribution \citep{Uttley2005}. This implies variability processes in BHXRB cannot be solely additive and must include multiplicative components which couple phenomena on many timescales. 

The bispectrum---along with the bicoherence derived from its amplitude and the biphase derived from its phase---has been used to detect such couplings \citep{Maccarone2002}. 
The bispectrum is a third-order Fourier construction which can be used to detect nonlinear couplings between frequencies in a single light curve \citep{Kim1979}. 
From many BHXRB---GX~339$-$4 \citep{Arur2019}, GRS~1915$+$105 \citep{Maccarone2011}, and Swift~J1727.8$-$1613 \citep{Zhu2024b}, among others---a consistent phenomenology of the bicoherence has been built. 
First named by \cite{Maccarone2011}, the ``cross'' state describes epochs where strong coupling is observed between the frequency of the QPO $(\nz)$ and a generic second frequency, which lie along the arms of the cross at $f_1 = \nz$ and $f_2 = \nz$; the ``hypotenuse'' state describes epochs with strong coupling between pairs of frequencies that sum to $\nz$, which lie along the ridge of the hypotenuse at $f_1 + f_2 = \nz$; and the ``web'' state describes epochs where both behaviors are observed. We also consider ``point'' states, where the bicoherence is high only at the harmonic point $(\nz,\nz)$, which is common in type B QPOs \citep{Arur2019,Arur2020}.
Finally, as described in \cite{Arur2020}, these states are correlated with $\nz$ and the inclination of the BHXRB\@. At low $\nz$, sources typically show the web state. As $\nz$ increases, high-inclination sources preferentially show the cross state and low-inclination sources preferentially show the hypotenuse state.

In this paper, we consider how the bispectrum manifests for a light curve model consisting of a lognormal accretion flow modulated by a QPO and its harmonic, a quadratically coupled term, and incoherent dilution. In Section~\ref{sec:math}, we review our mathematical conventions and the definition of the bispectrum. In Section~\ref{sec:model}, we describe the model we consider. In Sections~\ref{sec:bispec_derivation} and~\ref{sec:bico}, we show that the bispectrum of our model matches the observed bicoherence states. In Section~\ref{sec:transitions}, we present a pathway for a BHXRB to transition between those states as it evolves through outburst. In Section~\ref{sec:inclination}, we connect this evolution to the inclination of the source and offer an explanation for the observed correlation. Finally, we verify our description with numerical simulations in Section~\ref{sec:simulation} and discuss the implications and limitations of our model in Section~\ref{sec:discussion}.

During preparation we became aware of \cite{Jin2026inprep} who independently consider a similar idea. As we do, they consider the bispectrum, bicoherence, and biphase of broadband noise multiplicatively modulated by a QPO and its harmonic, similarly finding it produces bicoherence on the frequencies of the web state with opposite-sign biphase on the arms of the cross and the ridge of the hypotenuse. Unlike this work, they consider how this pattern manifests in observations, applying the modulus and biphase of the bispectrum to reveal patterns that are weak or difficult to distinguish in the bicoherence in Swift~J1727.8$-$1613 and MAXI~J1820$+$070. Unlike \cite{Jin2026inprep}, we explicitly separate the coupling between the QPO and the quadratic content of the broadband noise. We use this additional coupling to explain how transitions between bicoherence states could be driven and their possible connection to inclination.

\section{Mathematical Background and Notation}\label{sec:math}
We consider a light curve $x(t)$ with Fourier transform $X(f)$. The time series $x(t)$ is real and has Hermitian-symmetric Fourier transform, such that $X(f) = X^*(-f)$. The power spectrum of a light curve describes the amplitude of variability at a frequency $f$ in that light curve,
\begin{equation} \label{eq:power_spectrum}
    \mathcal{P}(f) = E[X(f) \; X^*(f)],
\end{equation}
where $E[\cdot]$ denotes an expectation value and $\cdot^*$ denotes a complex conjugation. In practice, we have access to only finite-duration time series. As such, expectation values are typically approximated with averaging over $K$ segments.

\subsection{Bispectrum}
The bispectrum is a higher-order Fourier technique used to detect nonlinear coupling between different frequencies in a time series. It is defined as
\begin{equation}\label{eq:bispectrum}
    B(f_1,f_2) = E[X(f_1) X(f_2)X^*(f_1 + f_2)].
\end{equation}
Like the power spectrum, the expectation value in the definition of the bispectrum is, in practice, approximated by averaging over $K$ segments. The resulting estimate of the bicoherence (Equation~\eqref{eq:bicoherence}) carries a bias of approximately $1/K$, which is routinely subtracted \citep[e.g.,][]{Arur2019,Arur2020,Arur2022}.

The bispectrum is a complex quantity and we can treat its amplitude and phase separately. The normalized squared amplitude of the bispectrum is known as the bicoherence \citep{Kim1979},\footnote{The original definition of \cite{Kim1979} defines the bicoherence as $b$ and calls $b^2$ the squared bicoherence. Later work \citep[notably,][]{Maccarone2002} calls $b^2$ the bicoherence. We define the bicoherence as $b^2$ for consistency.}
\begin{equation}\label{eq:bicoherence}
    b^2(f_1,f_2) = \frac{|B(f_1,f_2)|^2}{E[|X(f_1) X(f_2)|^2] E[|X(f_1+f_2)|^2]}.
\end{equation}
The bicoherence is bounded $0 \leq b^2 \leq 1$ and gives the fraction of power at $f_1 + f_2$ due to the coupling between the waves of frequencies $f_1$, $f_2$, and $f_1 + f_2$ \citep{Kim1979}.

The phase of the bispectrum is known as the biphase,
\begin{equation}\label{eq:biphase}
    \beta(f_1,f_2) = \arg[B(f_1,f_2)],
\end{equation}
chosen such that $\beta \in (-\pi,\pi]$. 

\subsection{Gaussian Moments}
Throughout, we utilize the following properties of the moments of zero-centered standard Gaussian noise, collected for convenience. Consider a real-valued time series $x(t)$ with Gaussian-distributed $X(f) = \gh(f) \sim \mathcal{N}(0,1)$. Then,
\begin{equation}\label{eq:moments}
    \begin{aligned}
    E[\gh(f)] &= 0\\
    E[\gh(f) \,\gh^*(f')] &\propto \delta_{f,f'}\\
    E[\gh(f) \,\gh(f')] &\propto \delta_{-f,f'}\\
    E[\gh(f_1) \,\gh(f_2) \,\gh(f_3)] &= 0.
    \end{aligned}
\end{equation}

\section{Light Curve Model}\label{sec:model}
We consider the bispectrum of the following light curve model:
\begin{equation}\label{eq:lc_model}
    x(t) = \underbrace{C \ell(t) M(t)}_{\text{Modulated flow}} + 
    \underbrace{x_Q(t)}_{\text{Quadratic Coupling}} + 
    \underbrace{D \left [ 1 + d(t)\right ]}_{\text{Dilution}}.
\end{equation}
The first term represents the inner accretion flow of the BHXRB modulated by the QPO: $\ell(t)  \equiv \exp \left [ g(t) - \frac{1}{2} \sigma_g^2\right ]$ is lognormal noise originating in the hot inner accretion flow \citep{Uttley2005, Ingram2011}. It is built from a Gaussian field $g(t)$ with variance $\sigma_g^2$, Fourier transform $\gh(f) \equiv N(f)$, and power spectrum $E[|N(f)|^2] \equiv P_g(f)$. We represent the QPO and its second harmonic with $M(t) = 1 + \sqrt{2} \; a_1 \cos \varphi + \sqrt{2} \; a_2 \cos (2 \varphi + \vartheta)$ with $\varphi \equiv 2 \pi \nu_0 t + \psi$. We treat $\psi$ as constant in a given light curve segment, but random segment-to-segment. 
For simplicity, we treat the QPO and its harmonic as pure tones.

We next add a component representing how the QPO couples to the quadratic power in the flow. We define $x_Q(t) \equiv K_2 (\mathcal{G} * q_2)(t)$, where $q_2(t) =  g^2(t) - \left \langle g^2(t) \right \rangle$, $\mathcal{G}(t)$ is a real kernel with Fourier transform $G(f) = G^*(-f)$, and $|G(\nz)| = 1$. With some positive real amplitude $K_2$, the QPO couples to the squared noise field $g^2(t)$ through some kernel $G$. We discuss the nature of this term in Section~\ref{sec:quadratic_nature}.

Finally, we consider all other sources of variability power not multiplicatively coupled to the inner flow and QPO: $D \left [ 1 + d(t)\right ]$. This dilution could be variability originating in the disk, the effects of an outflowing wind, or the effects of a changing obscuring column due to the wind of the companion star. We also absorb the contribution of the constant Poisson power floor in this term. 

\section{Model Bispectrum}\label{sec:bispec_derivation}
The Fourier transform of Equation~\eqref{eq:lc_model}, keeping the modulated flow term to linear order in the flow noise, is
\begin{equation} \label{eq:fourier}
\begin{aligned}
    X(f) \;=\;\; & C\Bigl[\; \delta(f) + N(f) \\
    & \hphantom{C\Bigl[\;}{}+ \tfrac{a_1}{\sqrt{2}}\bigl(e^{i\psi}\delta(f-\nu_0) + e^{-i\psi}\delta(f+\nu_0)\bigr) \\
    & \hphantom{C\Bigl[\;}{}+ \tfrac{a_1}{\sqrt{2}}\bigl(e^{i\psi}N(f-\nu_0) + e^{-i\psi}N(f+\nu_0)\bigr) \\
    & \hphantom{C\Bigl[\;}{}+ \tfrac{a_2}{\sqrt{2}}\bigl(e^{i(2\psi+\vartheta)}\delta(f-2\nu_0) + \\
    & \hphantom{C\Bigl[\;}{}e^{-i(2\psi+\vartheta)}\delta(f+2\nu_0)\bigr) \\
    & \hphantom{C\Bigl[\;}{}+ \tfrac{a_2}{\sqrt{2}}\bigl(e^{i(2\psi+\vartheta)}N(f-2\nu_0) + \\
    & \hphantom{C\Bigl[\;}{}e^{-i(2\psi+\vartheta)}N(f+2\nu_0)\bigr)\Bigr] \\
    & {}+ K_2\,G(f)\,\hat{q}_2(f) \\
    & {}+ D\bigl[\,\delta(f) + \hat{d}(f)\,\bigr].
\end{aligned}
\end{equation}
We are interested in triplets of frequencies $(f_1,f_2,f_1+f_2)$ where the bispectrum $B(f_1,f_2)$ is nonzero. That is, where $E[X(f_1) X(f_2) X^*(f_1+f_2)]$ is nonzero. This requires the phase $\Phi = \arg[X(f_1)] + \arg[X(f_2)] -\arg[X(f_1 + f_2)]$ to be constant between segments. For this reason, only terms for which all factors of $e^{i \psi}$ cancel will be considered. 

In the following subsections, we describe how the terms of Equation~\eqref{eq:fourier} multiply and couple together in the bispectrum at the frequencies of interest. In each, we consider how two sets of terms multiply at frequencies corresponding to the arms, the ridge, or the harmonic point. Outside these frequencies (or harmonics thereof) the products average to zero trivially. In each subsection, we consider only the terms of Equation~\eqref{eq:fourier} which correspond to both the coupling under study and which will produce a nonzero bispectrum. The results of this procedure are summarized in Table~\ref{tab:contributions}.

\subsection{Fundamental Bilinear Coupling}
We first turn to the terms of Equation~\eqref{eq:fourier} where the flow noise sidebands couple to the QPO fundamental. Let us take the frequency triple $(f_1, \nz, f_1 + \nz)$ with $f_1 < \nz$, corresponding to the arms. From Equation~\eqref{eq:fourier}, we take
\begin{equation} \label{eq:bilinear_cross_fouriers}
\begin{aligned}
    X(f_1) &\supset C \Bigl[N(f_1) \\
    &\quad {}+ \frac{a_1}{\sqrt{2}} \bigl(e^{i\psi}N(f_1-\nz) + e^{-i\psi} N(f_1+\nz)\bigr)\Bigr]\\
    X(\nz) &\supset C \Bigl[N(\nz) \\
    &\quad {}+ \frac{a_1}{\sqrt{2}} \bigl(e^{i\psi} + e^{i\psi} N(0) + e^{-i \psi} N(2 \nz)\bigr)\Bigr] \\
    X(f_1 + \nz) &\supset C \Bigl[N(f_1 + \nz) \\
    &\quad {}+ \frac{a_1}{\sqrt{2}} \bigl(e^{i\psi}N(f_1) + e^{-i\psi} N(f_1+2\nz)\bigr)\Bigr],
\end{aligned}
\end{equation}
then the bispectrum is 
\begin{equation} \label{eq:bilinear_cross_bispec}
    B^{\rm M}(f_1, \nz) = C^3 \frac{a_1^2}{2} \left [P_g(f_1) + P_g(f_1 + \nz)\right ],
\end{equation}
where we have applied the Gaussian moments of Equation~\eqref{eq:moments}, kept only terms without factors of $e^{i \psi}$, and used  $N(-f) = N^*(f)$ and $P_g(0) = 0$.

We next take the bilinear frequency triple $(f_1,f_2,\nz)$. Taking $X(f_1)$, $X(f_2)$, and $X(\nz)$ as in Equation~\eqref{eq:bilinear_cross_fouriers}, $f_1 + f_2 = \nz$, and calculating the bispectrum, we can write:
\begin{equation} \label{eq:bilinear_hyp_bispec}
    B^{\rm M}(f_1,f_2) = C^3 \frac{a_1^2}{2} \left [ P_g(f_1) + P_g(f_2) \right].
\end{equation}

\subsection{Harmonic Bilinear Coupling}
We next turn to the terms of Equation~\eqref{eq:fourier} where the flow noise sidebands couple to the QPO harmonic. Returning to the triple $(f_1,\nz,f_1+\nz)$ on the arms, the harmonic sidebands add
\begin{equation} \label{eq:harmonic_cross_fouriers}
\begin{aligned}
    X(f_1) &\supset C \frac{a_2}{\sqrt{2}} \bigl(e^{i(2\psi+\vartheta)}N^*(2\nz - f_1) \\
    & \hphantom{\supset C \frac{a_2}{\sqrt{2}} \bigl(}{}+ e^{-i(2\psi+\vartheta)} N(f_1+2\nz)\bigr)\\
    X(\nz) &\supset C \frac{a_2}{\sqrt{2}} \bigl(e^{i(2\psi+\vartheta)}N^*(\nz) \\
    & \hphantom{\supset C \frac{a_2}{\sqrt{2}} \bigl(}{}+ e^{-i(2\psi+\vartheta)} N(3\nz)\bigr)\\
    X(f_1+\nz) &\supset C \frac{a_2}{\sqrt{2}} \bigl(e^{i(2\psi+\vartheta)}N^*(\nz - f_1) \\
    & \hphantom{\supset C \frac{a_2}{\sqrt{2}} \bigl(}{}+ e^{-i(2\psi+\vartheta)} N(f_1+3\nz)\bigr)
\end{aligned}
\end{equation}
to the legs of Equation~\eqref{eq:bilinear_cross_fouriers}. Keeping only the terms of the product which contain at least one harmonic sideband, the bispectrum is
\begin{equation} \label{eq:harmonic_cross_bispec}
    B^{\rm R}(f_1,\nz) = C^3 \frac{a_1^2 a_2}{2\sqrt{2}}\, e^{-i\vartheta} \left[P_g(\nz - f_1) + P_g(f_1 + 2\nz)\right],
\end{equation}
where we have used the simplifications of Equation~\eqref{eq:bilinear_cross_bispec}. In addition, the harmonic offset $\vartheta$ is measured relative to $\psi$ and is therefore the same in every segment, so the factor $e^{-i\vartheta}$ survives averaging. 

On the ridge, with triple $(f_1,f_2,\nz)$ and $f_1+f_2=\nz$, the harmonic sidebands of $X(f_1)$ and $X(f_2)$ become
\begin{equation} \label{eq:harmonic_hyp_fouriers}
\begin{aligned}
    X(f_1) &\supset C \frac{a_2}{\sqrt{2}} \bigl(e^{i(2\psi+\vartheta)}N^*(f_2 + \nz) \\
    & \hphantom{\supset C \frac{a_2}{\sqrt{2}} \bigl(}{}+ e^{-i(2\psi+\vartheta)} N(f_1+2\nz)\bigr)\\
    X(f_2) &\supset C \frac{a_2}{\sqrt{2}} \bigl(e^{i(2\psi+\vartheta)}N^*(f_1 + \nz) \\
    & \hphantom{\supset C \frac{a_2}{\sqrt{2}} \bigl(}{}+ e^{-i(2\psi+\vartheta)} N(f_2+2\nz)\bigr),
\end{aligned}
\end{equation}
where we have used $f_1 - 2\nz = -(f_2+\nz)$, and $X(\nz)$ is as in Equation~\eqref{eq:harmonic_cross_fouriers}. The bispectrum is
\begin{equation} \label{eq:harmonic_hyp_bispec}
    B^{\rm R}(f_1,f_2) = C^3 \frac{a_1^2 a_2}{2\sqrt{2}}\, e^{i\vartheta} \left[P_g(f_1 + \nz) + P_g(f_2 + \nz)\right].
\end{equation}
The sign of the biphase is reversed relative to Equation~\eqref{eq:harmonic_cross_bispec} because the fundamental tone now enters through the conjugated leg $X^*(\nz)$.

\subsection{Quadratic Coupling}
We now consider the quadratic term $K_2 G(f) \hat{q}_2(f)$, where $\hat{q}_2(f) = \int N(\nu')\,N(f-\nu')\,d\nu'$ is the self-convolution of the flow noise. On the ridge $(f_1,f_2,\nz)$ we take
\begin{equation} \label{eq:quad_hyp_fouriers}
\begin{aligned}
    X(f_1) &\supset C N(f_1)\\
    X(f_2) &\supset C N(f_2)\\
    X(\nz) &\supset K_2 G(\nz) \int N(\nu')\,N(\nz-\nu')\,d\nu' \\
    &\supset 2 K_2 G(\nz) N(f_1) N(f_2),
\end{aligned}
\end{equation}
where the factor of 2 counts the two frequencies $\nu' = f_1$ and $\nu' = f_2$ in the integral. The bispectrum is
\begin{equation} \label{eq:quad_hyp_bispec}
    B^{\rm Q}(f_1,f_2) = 2\,C^2 K_2 G^*(\nz)\, P_g(f_1)\, P_g(f_2),
\end{equation}
where we have used $E[|N(f_1)|^2 |N(f_2)|^2] = P_g(f_1) P_g(f_2)$ for distinct Gaussian modes.

On the arms $(f_1,\nz,f_1+\nz)$, the integral instead contributes a difference pair from $\nu' = -f_1$ and $\nu' = f_1 + \nz$,
\begin{equation} \label{eq:quad_cross_fouriers}
\begin{aligned}
    X(f_1) &\supset C N(f_1)\\
    X(\nz) &\supset 2 K_2 G(\nz) N^*(f_1) N(f_1+\nz)\\
    X(f_1+\nz) &\supset C N(f_1+\nz),
\end{aligned}
\end{equation}
and the bispectrum is
\begin{equation} \label{eq:quad_cross_bispec}
    B^{\rm Q}(f_1,\nz) = 2\,C^2 K_2 G(\nz)\, P_g(f_1)\, P_g(f_1+\nz).
\end{equation}
Here $\hat{x}_Q$ enters unconjugated, so the biphase $+\alpha$, where $\alpha \equiv \arg[G(\nz)]$, has the opposite sign to that on the ridge. In both triples we have neglected the quadratic term on the other two legs, which carry $G(f_1)$, $G(f_2)$, or $G(f_1+\nz)$ and are small for a kernel peaked at $\nz$.

\subsection{Harmonic Coupling}
Finally, we consider the harmonic point $(\nz,\nz,2\nz)$, where the fundamental and harmonic couple directly. From Equation~\eqref{eq:fourier}, we take
\begin{equation} \label{eq:harmonic_point_fouriers}
\begin{aligned}
    X(\nz) &\supset C \Bigl[N(\nz) + \frac{a_1}{\sqrt{2}}\bigl(e^{i\psi} + e^{-i\psi}N(2\nz)\bigr) \\
    & \hphantom{\supset C \Bigl[}{}+ \frac{a_2}{\sqrt{2}}\bigl(e^{i(2\psi+\vartheta)}N^*(\nz) \\
    & \hphantom{\supset C \Bigl[{}+ \frac{a_2}{\sqrt{2}}\bigl(}{}+ e^{-i(2\psi+\vartheta)}N(3\nz)\bigr)\Bigr]\\
    X(2\nz) &\supset C \Bigl[N(2\nz) + \frac{a_1}{\sqrt{2}}\bigl(e^{i\psi}N(\nz) + e^{-i\psi}N(3\nz)\bigr) \\
    & \hphantom{\supset C \Bigl[}{}+ \frac{a_2}{\sqrt{2}}\,e^{i(2\psi+\vartheta)}\Bigr],
\end{aligned}
\end{equation}
so that the bispectrum $E[X(\nz)^2 X^*(2\nz)]$ is
\begin{equation} \label{eq:harmonic_point_bispec}
\begin{aligned}
    B^{\rm H}(\nz,\nz) = C^3 \Bigl[& \frac{a_1^2 a_2}{2\sqrt{2}}\, e^{-i\vartheta} \bigl(1 + 2 P_g(3\nz)\bigr) \\
    & + \bigl(a_1^2 + a_2^2\bigr) P_g(\nz) + a_1^2 P_g(2\nz)\Bigr].
\end{aligned}
\end{equation}
We have dropped the sideband terms with $N(0)$, since $P_g(0) = 0$, and used the same Gaussian moments as above. 

\begin{table*}
\centering
\caption{Bispectrum contributions that survive averaging, grouped by region.}
\label{tab:contributions}
\renewcommand{\arraystretch}{1.5}
\begin{tabular}{llll}
\hline
\hline
Origin & Frequency triple & Amplitude scaling & Biphase \\
\hline
\multicolumn{4}{l}{\textit{Arms of the cross}} \\
Bilinear fundamental $(B^{\rm M})$ & $(f_1,\,\nu_0,\,f_1+\nu_0)$ & $C^3\,\frac{a_1^2}{2}\bigl[P_g(f_1)+P_g(f_1+\nu_0)\bigr]$ & $0$ \\
Bilinear harmonic $(B^{\rm R})$ & $(f_1,\,\nu_0,\,f_1+\nu_0)$ & $C^3\,\frac{a_1^2 a_2}{2\sqrt{2}}\bigl[P_g(\nu_0-f_1)+P_g(f_1+2\nu_0)\bigr]$ & $-\vartheta$ \\
Quadratic $(B^{\rm Q})$& $(f_1,\,\nu_0,\,f_1+\nu_0)$ & $2\,C^2\,K_2|G(\nu_0)|\,P_g(f_1)\,P_g(f_1+\nu_0)$ & $+\alpha$ \\
\hline
\multicolumn{4}{l}{\textit{Ridge of the hypotenuse}} \\
Bilinear fundamental $(B^{\rm M})$ & $(f_1,\,f_2,\,\nu_0)$ & $C^3\,\frac{a_1^2}{2}\bigl[P_g(f_1)+P_g(f_2)\bigr]$ & $0$ \\
Bilinear harmonic $(B^{\rm R})$& $(f_1,\,f_2,\,\nu_0)$ & $C^3\,\frac{a_1^2 a_2}{2\sqrt{2}}\bigl[P_g(f_1+\nu_0)+P_g(f_2+\nu_0)\bigr]$ & $+\vartheta$ \\
Quadratic $(B^{\rm Q})$& $(f_1,\,f_2,\,\nu_0)$ & $2\,C^2\,K_2|G(\nu_0)|\,P_g(f_1)\,P_g(f_2)$ & $-\alpha$ \\
\hline
\multicolumn{4}{l}{\textit{Harmonic point}} \\
Tone--harmonic beat $(B^{\rm H})$& $(\nu_0,\,\nu_0,\,2\nu_0)$ & $C^3\,\frac{a_1^2 a_2}{2\sqrt{2}}$ & $-\vartheta$ \\
Sideband--tone beat at $3\nu_0$ $(B^{\rm H})$& $(\nu_0,\,\nu_0,\,2\nu_0)$ & $C^3\,\frac{a_1^2 a_2}{\sqrt{2}}\,P_g(3\nu_0)$ & $-\vartheta$ \\
Bilinear fundamental $(B^{\rm M})$& $(\nu_0,\,\nu_0,\,2\nu_0)$ & $C^3\,a_1^2\bigl[P_g(\nu_0)+P_g(2\nu_0)\bigr]$ & $0$ \\
Harmonic-sideband bias $(B^{\rm H})$& $(\nu_0,\,\nu_0,\,2\nu_0)$ & $C^3\,a_2^2\,P_g(\nu_0)$ & $0$ \\
\hline
\end{tabular}
\end{table*}

\section{Model Bicoherence}\label{sec:bico}
Having derived the bispectrum of the flow-QPO couplings present in our model, we now consider how these bispectral amplitudes manifest in the bicoherence, which normalizes the squared amplitude of the bispectrum. 
For weakly coupled terms, we can approximate the bicoherence\footnote{Equation~\eqref{eq:bico_approximation} is sometimes known as the ``skewness function''. The skewness and the bicoherence typically take similar values and are sometimes confused \citep[e.g., in][]{Proakis1992}. Notably, the skewness function is not correctly normalized such that it is bounded between 0 and 1. For a comprehensive overview, see \cite{Fackrell1997}.}
\begin{equation}\label{eq:bico_approximation}
    b^2(f_1,f_2) \approx \frac{\left | B(f_1,f_2)\right |^2}{\mathcal{P}(f_1) \mathcal{P}(f_2) \mathcal{P}(f_1 + f_2)},
\end{equation}
where $\mathcal{P}(f) \equiv E[X(f) X^*(f)]$ is the total power at a frequency $f$. This approximation neglects covariance between $|X(f_1)|^2$ and $|X(f_2)|^2$ due to the coupling discussed in Section~\ref{sec:bispec_derivation}. 
Corrections are expected to be at most a few percent.

Using Equation~\eqref{eq:bico_approximation} and the results of Section~\ref{sec:bispec_derivation}, we now calculate the bicoherence on the arms, the ridge, and the harmonic point. Define
\begin{equation}\label{eq:fractions}
\begin{aligned}
    \rho(f_a,f_b) &\equiv \frac{C^2 P_g(f_a)}{\Pow(f_b)}\\
    \tau &\equiv \frac{C^2 a_1^2/2}{\Pow(\nz)}\\
    \tau_2 &\equiv \frac{C^2 a_2^2/2}{\Pow(2\nz)}\\
    \Lambda &\equiv \frac{K_2^2 |G(\nz)|^2 P_{g^2}(\nz)}{\Pow(\nz)}
\end{aligned}
\end{equation}
as, respectively, the coupled ratio, the tone fraction, the harmonic tone fraction, and the quadratic fraction.
We have additionally denoted $P_{g^2}(\nu) \equiv 2\sum_{\nu'} P_g(\nu') P_g(\nu-\nu')$, twice the autoconvolution of the flow noise power spectrum. Because $g$ is Gaussian, Isserlis' theorem makes this the power spectrum of $g^2(t)$.
Each of these quantities measures the power located in a coupled mode compared to the total power. $\rho(f_a,f_b)$ measures the power in the flow noise at $f_a$ as a fraction of the total power at $f_b$; $\tau$ measures the fraction of power at $\nz$ in the QPO tone; $\tau_2$ measures the fraction of power at $2\nz$ in the harmonic tone; and $\Lambda$ measures the fraction of power in the locked quadratic mode.

The total power at the QPO frequency $\nz$ will be the sum of the power due to the QPO, the power due to the locked quadratic mode, the power due to the flow noise at $\nz$, and any diluting power. If we let the fraction of power at $\nz$ due to the flow noise and dilution be $\epsilon$, we can write
\begin{equation} \label{eq:qpo_budget}
    1 = \Lambda + \tau + \epsilon.
\end{equation}
We will assume the locked quadratic mode and QPO are strong, strong enough to treat the other sources of power as negligible at this frequency $(\epsilon \rightarrow 0)$. The flow noise and dilution are broadband noise, spanning many decades of frequency, but not powerful at any particular one. 

We continue by defining
\begin{equation}\label{eq:spm}
    s_{\pm}(f) \equiv \frac{a_1^2}{2} \rho(f,\nz\pm f),
\end{equation}
as the share of power at the image frequency $\nz \pm f$ that is the QPO's copy of the mode at $f$.

We additionally define 
\begin{equation}\label{eq:flow_ratio}
    r(f_a,f_b) = \frac{P_g(f_a)}{P_g(f_b)}
\end{equation}
as the ratio between frequencies in the flow noise. For red flow noise, $r(f_a>f_b,f_b) < 1$.
Finally, for compactness, we denote the special cases,
\begin{equation}\label{eq:rpm}
    r_{\pm}(f) = \frac{P_g(\nz \pm f)}{P_g(f)}
\end{equation}
and 
\begin{equation}\label{eq:short_rho}
    \rho(f) = \rho(f,f).
\end{equation}

\subsection{Bicoherence on the Arms of the Cross}\label{sec:cross_bico}
We can calculate the bicoherence on the arms, at frequency triplets $(f_1,\nz,f_1+\nz)$, by calculating the bicoherence of each surviving contributor (see Table~\ref{tab:contributions}) and summing using phasor addition. Separate contributors to the bicoherence with different biphases interfere with each other, meaning in general $b^2_{\rm total} \leq \bigl(\sum_i b_i\bigr)^2$.

Let $f_3 = f_1 + \nz$. Both bilinear bispectra on the arms ($B^{\rm M}$ and $B^{\rm R}$) are sums of flow-noise power at two frequencies. For $b^2_M$ we factor out the lower-frequency term, and for $b^2_R$ we factor out $P_g(f_1 + 2\nz)$; in each case the other term is written relative to the factored one. We find
\begin{equation} \label{eq:cross_bico_components}
\begin{aligned}
    b^2_M(f_1) &= \tau \, \rho(f_1) s_+(f_1) [1 + r_+(f_1)]^2\\
    b^2_R(f_1) &= \tau \frac{a_1^2 a_2^2}{4}\, \rho(f_1 + 2\nz,f_1)\, \rho(f_1 + 2\nz,f_3)\times \\ &\bigl[1 + r(\nz - f_1,\, f_1 + 2\nz)\bigr]^2 \\
    b^2_Q(f_1) &= \Lambda  \rho(f_1) \rho(f_3) \frac{4 P_g(f_1) P_g(f_3)}{P_{g^2}(\nz)}
\end{aligned}
\end{equation}
and the total bicoherence on the arms is
\begin{equation}
    b^2_{\rm cross} = \left | b_M + b_R \, e^{-i \vartheta} + b_Q \, e^{i \alpha}\right|^2.
\end{equation}

The bicoherence on the arms is dominated by the bilinear fundamental modulation term ($b^2_{M}$). The bicoherence due to the bilinear harmonic modulation contributes little in amplitude and matters most near $f_1 \approx \nz$, where it walks the biphase measured on the arms toward $-\vartheta$: $b^2_R/b^2_M \sim \frac{a_2^2}{2} r^2(\nz - f_1,f_1)$. Meanwhile, the bicoherence due to the quadratic channel on the arms scales as $b^2_Q \propto P^2_g(f_3)$, so it will be small for reasonably red flow noise \citep[e.g., the zero-centered noise with width set by the viscous timescale at the outer radius of the hot inner flow considered by][]{Ingram2011, Ingram2013}. 

We then have $b^2_{\rm cross} \approx \tau \rho(f_1) s_+(f_1)$. Notice the bicoherence is strongly dependent on the amplitude of the QPO, as $\tau, s_+ \propto a_1^2$. Low-frequency dilution enters this mode only once through $\rho(f_1)$.

\subsection{Bicoherence on the Ridge of the Hypotenuse}\label{sec:hyp_bico}
On the ridge, the frequency triplets are $(f_1,f_2,\nz)$ with $f_2 = \nz - f_1$. The same three channels survive here as on the arms (see Table~\ref{tab:contributions}). The bicoherence is symmetric under $f_1 \leftrightarrow f_2$, so we take $f_1 \le f_2$ and factor out the lower-frequency term of each bilinear sum.
We find
\begin{equation} \label{eq:hyp_bico_components}
\begin{aligned}
    b^2_M(f_1) &= \tau \rho(f_1) s_-(f_1) [1 + r_{-}(f_1)]^2 \\
    b^2_R(f_1) &= \tau \frac{a_1^2 a_2^2}{4}\, \rho(f_1 + \nz,f_1)\, \rho(f_1 + \nz,f_2)\times \\ &\bigl[1 + r(f_2 + \nz,f_1+\nz)\bigr]^2 \\
    b^2_Q(f_1) &= \Lambda \rho(f_1) \rho(f_2) \frac{4 P_g(f_1) P_g(f_2)}{P_{g^2}(\nz)}
\end{aligned}
\end{equation}
and the total bicoherence on the ridge is
\begin{equation}
    b^2_{\rm hyp} = \left | b_M + b_R \, e^{i \vartheta} + b_Q \, e^{-i \alpha}\right|^2.
\end{equation}

We find again that the bicoherence due to the bilinear harmonic modulation is subdominant here because it needs strong flow noise above $\nz$: $b^2_R/b^2_M \sim \frac{a_2^2}{2} r^2_+(f_1)$. We ignore its contributions for the same reason we can ignore $b^2_Q$ on the arms. 

In contrast with the bicoherence on the arms, two channels $b^2_M$ and $b^2_Q$ compete and interfere on the ridge. Neither coupling mode is suppressed because both frequencies $f_1$ and $f_2$ on the ridge sit below the QPO. 
As on the arms, $b^2_M \propto a_1^4$. However, this is offset on the ridge by the quadratic mode $b^2_Q \propto \Lambda \sim 1-\tau$. Through this second channel, the bicoherence on the ridge can survive a reduction in QPO amplitude that would dampen the arms. 
On the other hand, the ridge is doubly sensitive to low-frequency dilution because $b_Q^2 \propto \rho(f_1) \rho(f_2)$ and $b_M^2 \propto \rho(f_1) s_-(f_1)$, all of which depend on the strength of the flow relative to the total power below the QPO. 

\subsection{Bicoherence at the Harmonic Point}\label{sec:point_bico}
At the harmonic point $(\nz,\nz,2\nz)$, two legs of the bispectrum share the bin at $\nz$, so the denominator of Equation~\eqref{eq:bicoherence} does not factor as in Equation~\eqref{eq:bico_approximation}. 
For a tone in Gaussian noise, $E[|X(\nz)|^4] = (2-\tau^2)\,\Pow(\nz)^2$. We use this in place of $\Pow(\nz)^2$, i.e., the approximation of Equation~\eqref{eq:bico_approximation} is not used here. 
Normalizing each of the four terms of Equation~\eqref{eq:harmonic_point_bispec} (see Table~\ref{tab:contributions}), we find
\begin{equation}
\begin{aligned}
    b^2_H &= \frac{\tau^2 \tau_2}{2-\tau^2} \\
    b^2_S &= \frac{a_1^4\, \tau_2}{2-\tau^2}\, \rho(3\nz,\nz)^2 \\
    b^2_M &= \frac{2 \tau a_1^2}{2-\tau^2} \Bigl[ \sqrt{\rho(\nz,\nz) \rho(\nz,2\nz)} \\
    &\qquad {}+ \sqrt{\rho(2\nz,\nz) \rho(2\nz,2\nz)}\,\Bigr]^2 \\
    b^2_B &= \frac{2 a_2^2\, \tau_2}{2-\tau^2}\, \rho(\nz,\nz)^2
\end{aligned}
\end{equation}
for the tone--harmonic beat, the sideband--tone beat at $3\nz$, the bilinear fundamental, and the harmonic-sideband bias, respectively. We again make the assumption that the QPO and harmonic dominate the flow noise at their central frequencies, meaning all $\rho(f,\nz) \approx 0$ and $\rho(f,2\nz) \approx 0$.
The total bicoherence at the harmonic point is then
\begin{equation}
    b^2_{\rm point} = \frac{\tau^2 \tau_2}{2-\tau^2}.
\end{equation}

\section{Transitions}\label{sec:transitions}
Perhaps more interesting than any individual bicoherence state is how the transitions between these states are driven. \cite{Maccarone2011} first observed that GRS~1915$+$105 shows the web, cross, and hypotenuse states in different observations with similar power spectra. \cite{Arur2020} observed correlations between bicoherence state and $\nz$ and inclination where the web is ubiquitous at low $\nz$ while at high $\nz$ sources at high inclination show the cross state more often and sources at low inclination show the hypotenuse state more often. The model developed here offers an explanation for these phenomena. 

We will consider how the bicoherence functions developed in Section~\ref{sec:bico} change as a BHXRB undergoes a transition from the hard state through the hard intermediate state, and into the soft intermediate state. We assume that the source exhibits a type C QPO at the beginning of outburst which rises in frequency and falls in amplitude as it softens and eventually transitions into a type B QPO in the soft intermediate state. We also assume the overall RMS variability of the source decreases as it softens and that the source becomes brighter and more disk-dominated. This evolution simplifies many complex phenomena present during outburst and neglects how BHXRB differ from each other \citep{Remillard2006}. We treat this evolutionary path as only an approximation.

\subsection{Controls}\label{sec:controls}
To parameterize the bicoherence transitions through the spectral evolution of the BHXRB, we define the following two dimensionless parameters which tune the balance between the quadratic and bilinear modes and the level of dilution at low frequencies. We will consider only frequencies $f_1 \leq \nz/2$. First, we define the coupling ratio:
\begin{equation}\label{eq:coupling_ratio}
\begin{aligned}
    \mathcal{Q}(f_1\leq \nz/2) &\equiv \left | \frac{B^{\rm Q}(f_1)}{B^{\rm M}(f_1)}\right |_{\rm hyp} = \left.  \sqrt{\frac{b^2_{Q}(f_1)}{b^2_M(f_1)}} \right |_{\rm hyp} \\
    &= \frac{4 K_2}{C a_1^2} \frac{P_g(f_1) P_g(f_2)}{P_g(f_1) + P_g(f_2)},
\end{aligned}
\end{equation}
as the strength of the quadratic coupling measured in units of the squared modulation depth. 
Secondly, we define the dilution fraction,
\begin{equation}\label{eq:dilution_fraction}
    \eta_\pm(f) \equiv \frac{D^2 P_d(\nz \pm f)}{\Pow_0(\nz \pm f)},
\end{equation}
where $P_d \equiv E[|\hat{d}|^2]$ is the power spectrum of the diluting noise and $\Pow_0$ is the undiluted power spectrum defined by $\Pow(f) = \Pow_0(f) + D^2 P_d(f) $. The dilution fraction is the incoherent power at frequency $\nz \pm f$ relative to all other power at that frequency. Then, using the simplified forms of the bicoherence on the arms and the ridge discussed in Section~\ref{sec:bico} (dropping $b^2_R$ on both and dropping $b^2_Q$ on the arms), we can write
\begin{equation}\label{eq:bico_ratio}
    \frac{b^2_{\rm hyp}(f_1)}{b^2_{\rm cross}(f_1)} \approx W(f_1) \frac{1 + \eta_+(f_1)}{1 + \eta_-(f_1)} \left | 1 + \mathcal{Q} e^{-i \alpha} \right |^2,
\end{equation}
where 
\begin{equation}
    W(f_1) \equiv \frac{\Pow_0(f_1+\nz)}{\Pow_0(\nz - f_1)} \left [\frac{1 + r_-(f_1)}{1+r_+(f_1)} \right ]^2 \sim O(1)\leq 4,
\end{equation}
where the final inequality holds for all red $\Pow_0$. 

\subsection{Hard states}\label{sec:hard}
First, consider a BHXRB in the bright hard state, exhibiting a type C QPO with low $\nz$ and low disk flux \citep[as is common;][]{Motta2011,Ingram2019}.
In this state, we assume the flow noise is dominant over incoherent dilution and the QPO is strong, i.e., $\eta_- \sim 0$, $\mathcal{Q} \ll 1$, and $\rho(f_1) \sim 1$. Thus, $b^2_{\rm hyp}/b^2_{\rm cross} \approx W \sim 1$ and both the arms and the ridge are lit and of approximately equal strength---the web state.

\subsection{Intermediate states}\label{sec:intermediate}

As the source softens, three effects apply pressure to the bicoherence on the arms and the ridge. First, the RMS variability of the coupled flow noise will decrease. This will dim both the ridge and the arms by lowering $\rho$ and $s_\pm$. However, the two regions may not fade at the same rate. 

At the same time, we expect the QPO amplitude to decrease and the dilution power to increase. The QPO amplitude decreases, increasing $\mathcal{Q} \propto a_1^{-2}$, shifting the balance toward the quadratic mode and preserving the ridge over the arms. The dilution power---e.g., originating in the disk, which rises in prominence as the source becomes more disk-dominated, or due to a wind launched as the source becomes more luminous---has the effect of increasing $\eta_-(f_1)$, diluting the bicoherence on the ridge. The balance between these effects is described in Equation~\eqref{eq:bico_ratio} and ultimately controls which of the hypotenuse or cross states is shown as the source softens. The QPO amplitude decreasing tilts toward the hypotenuse state, the low-frequency dilution increasing tilts toward the cross state, and a balance between the two retains the web state as the overall bicoherence decreases.

\subsection{Soft intermediate state}\label{sec:soft}
Finally, we consider the soft intermediate state and a type B QPO\@. Type B QPOs are not accompanied by strong broadband noise. As such, the flow noise power at all frequencies in this state will be near zero. All terms of the bicoherence on the arms and the ridge require flow noise power. In this state, both the arms and the ridge disappear. 

However, the bicoherence at the harmonic point $(\nz,\nz,2\nz)$ does not depend on the flow noise. The bicoherence there is controlled primarily by the amplitude of the QPO and its harmonic. For type B QPOs with a harmonic, we expect to retain strong bicoherence at the harmonic point. 

\subsection{Inclination dependence}\label{sec:inclination}

The behavior in intermediate states predicted by this model naturally explains the inclination dependence observed by \cite{Arur2020}. Whether an intermediate state shows the cross or hypotenuse state is controlled by Equation~\eqref{eq:bico_ratio} and the balance between the low-frequency dilution ($\eta_-(f_1)$) and the strength of the QPO (via the coupling ratio $\mathcal{Q}$). The amplitude of type C QPOs rises with inclination \citep{Motta2015,Heil2015}. As such, for high-inclination sources $\mathcal{Q}$ stays low and the arms stay detectable as $\eta_-$ rises, while the weaker QPOs of low-inclination sources raise $\mathcal{Q}$ and favor the hypotenuse state. Dilution should push in the same direction. Equatorial disk winds are observed commonly in high-inclination sources \citep{Ponti2012}.\footnote{\cite{Ponti2012} find disk winds are ubiquitous in high-inclination sources in the high soft state. Winds have been occasionally observed in BHXRB in the intermediate state in X-rays \citep[e.g.,][]{Parra2025} and in the optical \citep[e.g.,][]{Sanchez2024}. The evolution of winds through the hard to soft state transitions is not well understood \citep{Shidatsu2026}.} If the dilution is wind-driven and present during the intermediate states, $\eta_-$ is larger in these sources, and Equation~\eqref{eq:bico_ratio} moves them further toward the cross state.

\section{Simulated Verification} \label{sec:simulation}
We verify our results using numeric simulations. We directly simulate the light curve model of Equation~\eqref{eq:lc_model} and calculate the resultant bicoherence and observe the expected states and behaviors. The flow noise is a Gaussian field $g(t)$ with tunable RMS $\sigma_g$. We choose the power spectrum $P_g(f)$ and simulate a light curve via the method of \cite{TimmerKonig1995}, which is then exponentiated to ensure lognormality ($\ell(t)$ in Equation~\eqref{eq:lc_model}). It multiplies the QPO waveform $M(t)$. The phase offset of the QPO $\psi_i$ is drawn uniformly on $[0, 2\pi)$ for each segment, so the QPO phase is incoherent between segments.

We simulate the quadratic coupling term by taking the quadratic content of the field $q_2(t)$, Fourier transforming it, and multiplying by a kernel $G$ and coupling constant $K_2$. We simulate only the simplest quadratic kernel, a flat kernel
\begin{equation}
    G(f) = \begin{cases}
        e^{i\alpha\,{\rm sgn}(f)}, &|f| \in [\nz - \delta, \nz + \delta]\\
        0, &\text{otherwise},
    \end{cases}
\end{equation}
and transform the result back to the time domain. 

The final addition is the dilution term $D[1+ d(t)]$. Similar to the flow field, the power spectrum of $d(t)$, $P_d(f)$, is chosen, simulated using the method of \cite{TimmerKonig1995} and inverse-transformed to the time domain. We convert the resultant light curve to counts by drawing from the Poisson distribution in each time bin.

Each light curve is simulated with timestep $\Delta t = 1/64$~s and divided into $K = 1000$ segments of 2048 bins ($\Delta f = 1/32$~Hz). We subtract the mean of each segment. When calculating the bicoherence, we use the normalization of \cite{Kim1979} and subtract the Poisson noise bias in the bispectrum detailed by \cite{Wirnitzer1985}.

\subsection{Pattern simulation} \label{sec:pattern_simulation}
We simulate four representative configurations, one for each of the web, cross, hypotenuse, and point states. The flow noise power spectrum $P_g$ is a broken power law, which is flat below $1.5$~Hz and $P_g \propto f^{-2}$ above. The power spectrum is normalized to a fractional RMS amplitude of $\sigma_g = 0.2$. The QPO central frequency is set to $\nz = 1.5$~Hz, the fractional amplitudes of the fundamental and harmonic are, respectively, $a_1 = 0.2$ and $a_2 = 0.05$. The harmonic is offset in phase from the fundamental by $\vartheta = \pi/2$~radians. The quadratic coupling amplitude $K_2$ is set so that the quadratic term $x_Q$ has a fractional RMS of $0.1$, and the flat kernel has half-width $\delta = 0.1$~Hz and phase $\alpha = \pi/2$. Finally, we represent the dilution with another broken power law, flat below $1.5$~Hz and $P_d \propto f^{-2}$ above. The dilution is weak, with $D = 0.3$ and $\sigma_d = 0.03$, where $\sigma_d \equiv D\,\mathrm{RMS}[d(t)]$ is the RMS of the dilution term in units of the mean flow. Each light curve segment is simulated at a mean rate of $6000$~ct~s$^{-1}$. 

\subsubsection{Simulated Web}

This fiducial run (Figure~\ref{fig:patterns}, top row) produces the web state. The arms at $f_1,f_2 = \nz$ and the ridge at $f_1 + f_2 = \nz$ exhibit comparable, strong bicoherence. Fainter copies appear at $\nz \rightarrow 2 \nz$, where the harmonic modulates the flow. This is the regime of Section~\ref{sec:hard}, with $\eta_-(f_1) \approx 0$ ($\eta_-(\nz/2) = 0.053$) and $\mathcal{Q} \ll 1$ ($\mathcal{Q}(\nz/2) = 0.10$). 

In this state, the biphase on the ridge is approximately zero. The dominant bilinear fundamental mode has zero biphase, and the two subdominant modes interfere with the bilinear harmonic mode contributing $+\vartheta = \pi/2$ and the quadratic mode contributing $-\alpha = -\pi/2$ (see Table~\ref{tab:contributions}). In the cells immediately transverse to the ridge, the quadratic mode dominates. This is due to our simulations using a quadratic kernel of non-zero width but a pure QPO tone. Similarly, on the arms, interference between the bilinear harmonic and quadratic modes zeros the biphase. Transverse to the arms, the quadratic mode is able to show through.

\subsubsection{Simulated Cross}

To produce the cross state we modify the fiducial simulation to match the description of the transition in Section~\ref{sec:intermediate}. We raise the dilution power at low frequency by raising $D$ to $D = 3.5$ and move the bulk of the dilution to low frequencies by lowering the break frequency of $P_d$, below which it is flat, from $1.5$~Hz to $0.5$~Hz. This second run is shown in the second row of Figure~\ref{fig:patterns}. The arms are visible while the ridge has fallen mostly to the noise floor. The arms are affected by the dilution, as they are weaker than in the web state, but they are not as diminished as the ridge, as predicted in Section~\ref{sec:intermediate}. In this configuration, $\eta_-(\nz/2) = 4.01$ and $\mathcal{Q}(\nz/2) = 0.10$.

The biphase throughout this simulation is near zero, on account of the dominance of the bilinear fundamental mode, which has zero biphase. The exception is the harmonic point $(\nz,\nz)$ which has $\beta = -\vartheta = -\pi/2$, as predicted. 

\subsubsection{Simulated Hypotenuse}
To produce the hypotenuse state, we again start with the fiducial simulation and tune the parameters. We decrease the amplitude of the QPO fundamental to $a_1 = 0.01$ and raise the QPO central frequency to $\nz = 5.0$~Hz. In this run (Figure~\ref{fig:patterns} third row), the only strong bicoherence is on the ridge and the biphase is $\beta \approx -\pi/2$, indicating the quadratic channel is controlling. Here, $\eta_-(\nz/2) = 0.11$ and $\mathcal{Q}(\nz/2) = 25.41$.

\subsubsection{Simulated Point}
\begin{figure*}
    \centering
    \includegraphics[width=\linewidth]{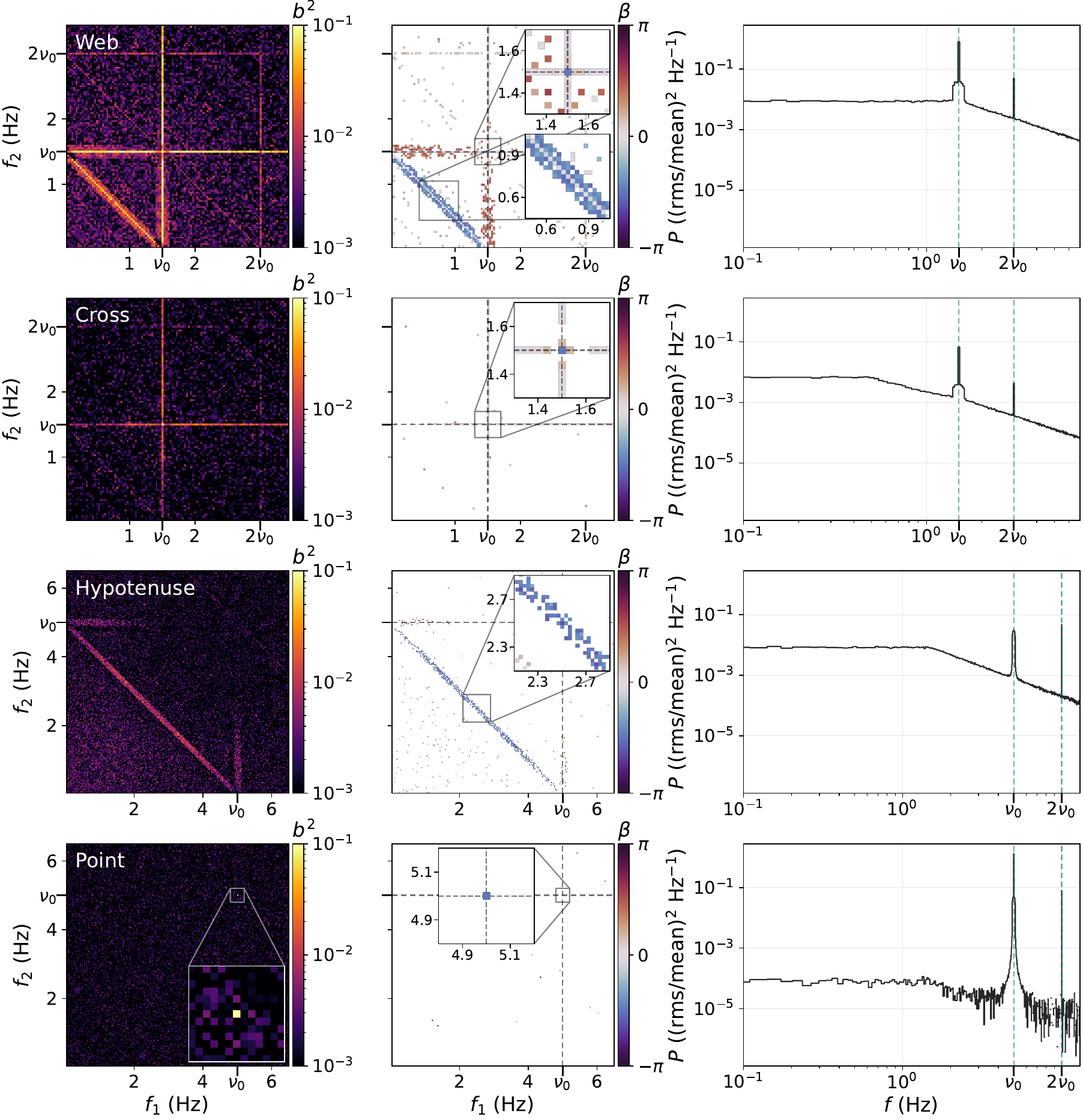}
    \caption{Bicoherence maps (left), biphase maps (middle), and power spectra (right) of the simulated light curves described in Section~\ref{sec:pattern_simulation}. The four configurations correspond to the (from top to bottom) web, cross, hypotenuse, and point states. In the biphase plots, we blank out the measured value and show white for all frequencies with bicoherence $b^2 < 8/K$---$8$ times the statistical floor set by the number of segments averaged---for visual clarity. We show the bicoherence with the Poisson noise bias of \cite{Wirnitzer1985} subtracted and the power spectrum with the constant Poisson noise floor subtracted. Where needed, insets are used to show areas of interest. In the second and third columns, thin dashed lines mark the QPO and harmonic frequencies ($\nz$ and $2\nz$).}
    \label{fig:patterns}
\end{figure*}

Finally, we simulate a type B QPO-like point state by starting with the fiducial parameters and lowering all variability below the QPO by setting $\sigma_g = 0.015$ and $D = 0.005$. We also increase the QPO frequency to $\nz = 5$~Hz, like a type B QPO\@. As expected, the only significant bicoherence here is at the harmonic point $(\nz,\nz)$. At all other frequency pairs, there is no flow noise to couple to. At the harmonic point, the biphase is set by the QPO harmonic offset $\beta = -\vartheta = -\pi/2$. This state is shown in the bottom row of Figure~\ref{fig:patterns}.

\subsection{Transition simulation} \label{sec:transition_sim}

We also use numerical simulations to verify the transition conditions described in Section~\ref{sec:transitions} and bicoherence ratio of Equation~\eqref{eq:bico_ratio}. We simulate many model light curves, each starting from a variant of the fiducial web simulation (Section~\ref{sec:pattern_simulation}) with $\nz = 5$~Hz, a steeper flow noise spectrum ($P_g \propto f^{-5}$ above the break), and dilution power truncated above $\nz$. We then tune $a_1$ and $\sigma_d$ to set $\mathcal{Q}$ and $\eta_-$, respectively. We simulate 31 different values of $a_1$, evenly spaced in log-space from 0.008 to 0.5. We simulate 36 different values of $\sigma_d$: $\sigma_d = 0$, six evenly log-spaced points below 0.01, and 29 evenly log-spaced points from 0.01 to 0.4. In total, we use 1,116 simulation runs, each with $K = 3000$ segments.

\begin{figure*}
    \centering
    \includegraphics[width=.7\linewidth]{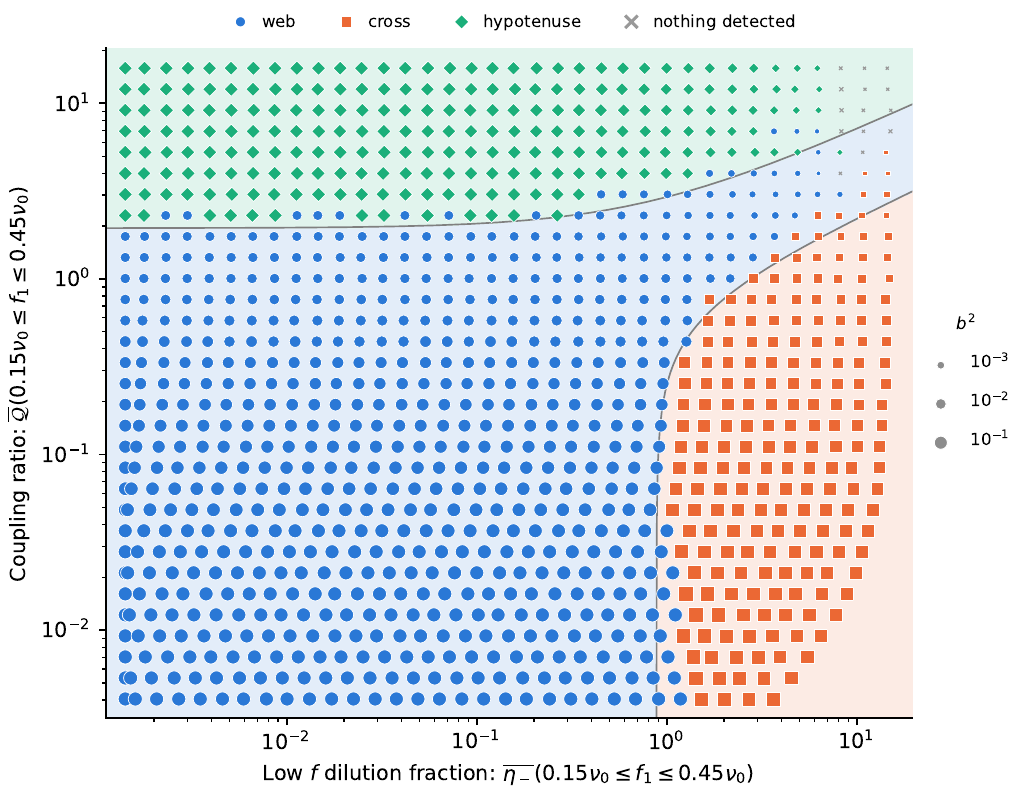}
    \caption{The observed bicoherence state for 1,116 simulated model light curves, simulated in the manner described in Section~\ref{sec:transition_sim}. We measure the bicoherence of each simulation and assign it a state.  We mark the observed state of each simulation at its observed dilution fraction and coupling ratio. The size of each marker corresponds to the mean bicoherence on the arms of the cross and the ridge of the hypotenuse and the color and shape correspond to the observed state. The background colors and contours are set by the predictions of Equation~\eqref{eq:bico_ratio}. Nondetections, where the bicoherence on the arms of the cross and the ridge of the hypotenuse are below 3 times the standard error, are marked with gray crosses.}
    \label{fig:state_grid}
\end{figure*}

For each simulation we measure $\eta_-$ and $\mathcal{Q}$ over the range $f_1 \in [0.15 \nz, 0.45 \nz]$ and place it on the grid. The state is predicted from these values using Equation~\eqref{eq:bico_ratio}. For these purposes, we define the cross state as those with $b^2_{\rm cross} \geq 3 b^2_{\rm hyp}$, hypotenuse states as those with $b^2_{\rm hyp} \geq 3 b^2_{\rm cross}$, and web states as all those in between. These definitions based on $\eta_-$ and $\mathcal{Q}$ define the contours and colored regions of Figure~\ref{fig:state_grid}. We then measure the bicoherence in the same frequency range on both the arms and the ridge and use the preceding definition to get an empirically observed state. When both bicoherences are below 3 standard errors from zero, we mark the simulation as a nondetection. These observed states make up the markers of Figure~\ref{fig:state_grid}. As expected, the web state is observed where $1 + \mathcal{Q}^2$ and $1 + \eta_-$ are comparable. When the coupling ratio grows, the web state gives way to the hypotenuse state and when the dilution fraction grows the web state gives way to the cross state. When both are high, the coupled power is strongly diluted and the QPO is weak and we do not detect bicoherence. We observe good agreement between our prediction and the measured state with $96\%$ of runs with detected bicoherence agreeing with predictions. 

\section{Discussion}\label{sec:discussion}

\subsection{The nature of the quadratic term}\label{sec:quadratic_nature}

The quadratic coupling term, $x_Q(t)$ in Equation~\eqref{eq:lc_model}, is a necessary part of the model because it allows a second channel to put bicoherence on the ridge. This unlinks the bicoherence on the ridge from that on the arms, so the ridge can stay lit when a low QPO amplitude has dimmed the arms. The quadratic coupling links the QPO to the quadratic content of the lognormal flow noise field. The bilinear mechanisms couple the QPO to only the linear content of the flow noise. 

We do not identify the process behind $x_Q$, but we do constrain its properties. The mode needs power at $\nz$ and must not depend on the QPO phase $\psi$ (or else it would average to zero). This second criterion excludes processes in which the QPO acts on the flow such as by modulating the disk truncation radius. We instead suggest this term could arise from a part of the QPO power which is driven by the noise.

Quadratic driving of the QPO is easy to come by. The flow noise is lognormal, so a torque that scales with the accretion rate, such as the accretion of misaligned angular momentum onto a precessing hot flow at the truncation radius \citep{Ingram2009, Ingram2019}, also carries the quadratic content $g^2$.

Further analysis of the bicoherence and the biphase along the ridge during hypotenuse states may be able to reveal the physical process(es) underlying this quadratic coupling. The bicoherence can probe the magnitude of this coupling and the biphase the phase. In particular, the biphase and bicoherence transverse to the ridge can measure the change in phase and magnitude of the kernel $G$ off of the QPO frequency. 

\subsection{Caveats and limitations}

Our analysis is not without limitations. First, the model of the light curve we use (Equation~\eqref{eq:lc_model}) makes several approximations. X-ray binary QPOs are merely quasi-periodic, not pure tones as we have used, i.e., they have some non-zero frequency width. Approximating the QPO and its harmonic as pure tones has allowed for the analytic coupling analysis of Sections~\ref{sec:bispec_derivation} and~\ref{sec:bico}. It is likely that replacing pure tones with quasi-periodic processes would smear out the resulting bicoherence patterns, giving the arms and the ridge a significant width. Additionally, we consider only the linear content of the lognormal flow, truncating the coupling at the first order. 

When deriving the bicoherence predicted by our model, we similarly make approximations to aid in developing analytical results. We have approximated the bicoherence with the skewness function, neglecting the covariance between $|X(f_1)|^2$ and $|X(f_2)|^2$. We then also ignored subdominant contributions to the bicoherence on the arms, the ridge, and the harmonic point. Finally, when deriving the coupling ratio and dilution fraction which parameterize the transitions between web, cross, hypotenuse, and point states we limit ourselves to only $f_1 \leq \nz/2$, as the contributions of the harmonic bilinear coupling mode become significant on the arms past this frequency.

However, many of these simplifications are not made when we numerically simulate our light curve model and verify it matches our predictions. We do not truncate the flow noise at linear order, we do not replace the bicoherence with skewness, and we do not ignore any contributors to the bicoherence. 

Finally, the bispectral properties of BHXRB are known to modulate with energy. For example, in Swift~J1727.8$-$1613 \cite{Zhu2024b} observe the bicoherence on the ridge strengthen with energy and  a ``parallel'' pattern emerge at frequencies $f_1 + f_2 = 2 \nz$ in the $10-20$ keV band. The coupling constants and amplitudes used in our model should thus be treated as energy-dependent.

\subsection{Conclusions}

A simple physically based model of an active black hole light curve is sufficient to create nonzero bicoherence at the frequencies corresponding to the patterns observed in BHXRB\@. We find that when a QPO multiplicatively modulates lognormal noise originating in an accretion flow, couplings between the QPO, its harmonic, and the flow noise produce high bicoherence at the frequencies of the web state commonly observed in BHXRB\@. When an additional quadratic coupling mode is present, the ridge of the hypotenuse and the arms of the cross that compose the web can exist independently. We find that the arms of the cross are dimmed by decreasing QPO amplitude and the ridge of the hypotenuse is dimmed by increasing the incoherent dilution at low frequencies. These transitions can explain why high-inclination sources, which have higher QPO amplitudes and may have stronger wind-driven dilution, show the cross state more often, while low-inclination sources, which have weaker QPOs, show the hypotenuse state.

Our model additionally predicts the biphase on the arms of the cross and the ridge of the hypotenuse. The biphase is controlled by interference between bilinear fundamental, bilinear harmonic, and quadratic coupling modes. The fundamental mode produces zero biphase, the harmonic mode produces biphase according to the phase offset of the harmonic, and the quadratic mode produces biphase according to the phase of the coupling kernel. These biphases have opposite signs on the ridge of the hypotenuse and the arms of the cross. 

This model provides the groundwork for a plausible physical explanation for the bicoherence states observed in black hole X-ray binaries. This work can provide a foundation for nonlinear variability content to probe the physical processes underlying accretion and the QPO. Such analysis could help to distinguish between models of the QPO which produce the same power spectral signatures. 

\section*{Data Availability}
Code to reproduce the numerical simulations and figures of this paper can be found publicly at \url{https://github.com/SethHtes/bispectrum-states-sims}.

\begin{acknowledgments}
    We disclose the use of GAI tools (Claude Opus 5 and 5.5, Anthropic) to write and refine code associated with this manuscript and to review the author's mathematical reasoning. We thank all members of the ``bispectrum support group'', especially Kavitha Arur and Tom Maccarone, for useful discussions. We thank Edward Nathan for pointing us to the Poisson noise correction in the bispectrum. SRL acknowledges funding from NASA grant 80NSSC24K0616. 
\end{acknowledgments}

\software{Jupyter \citep{2007CSE.....9c..21P,kluyver2016jupyter},
    Matplotlib \citep{Hunter:2007},
    Numpy \citep{numpy},
    Python \citep{python}
}

\bibliography{bib}{}

@article{Arur2019,
	title = {Nonlinear {Variability} of {Quasi}-{Periodic} {Oscillations} in {GX} 339-4},
	volume = {486},
	copyright = {https://academic.oup.com/journals/pages/open\_access/funder\_policies/chorus/standard\_publication\_model},
	issn = {0035-8711, 1365-2966},
	url = {https://academic.oup.com/mnras/article/486/3/3451/5454768},
	doi = {10.1093/mnras/stz1052},
	language = {en},
	number = {3},
	urldate = {2026-07-14},
	journal = {Monthly Notices of the Royal Astronomical Society},
	author = {Arur, K and Maccarone, T J},
	month = jul,
	year = {2019},
	pages = {3451--3458},
}

@article{Arur2020,
	title = {A likely inclination dependence in the non-linear variability of quasi-periodic oscillations from black hole binaries},
	volume = {491},
	copyright = {https://academic.oup.com/journals/pages/open\_access/funder\_policies/chorus/standard\_publication\_model},
	issn = {0035-8711, 1365-2966},
	url = {https://academic.oup.com/mnras/article/491/1/313/5583058},
	doi = {10.1093/mnras/stz2784},
	language = {en},
	number = {1},
	urldate = {2026-07-14},
	journal = {Monthly Notices of the Royal Astronomical Society},
	author = {Arur, K and Maccarone, T J},
	month = jan,
	year = {2020},
	pages = {313--323},
}

@article{Arur2022,
	title = {Using the bispectrum to probe radio {X}-ray correlations in {GRS} 1915+105},
	volume = {514},
	copyright = {https://academic.oup.com/journals/pages/open\_access/funder\_policies/chorus/standard\_publication\_model},
	issn = {0035-8711, 1365-2966},
	url = {https://academic.oup.com/mnras/article/514/2/1720/6595346},
	doi = {10.1093/mnras/stac1463},
	language = {en},
	number = {2},
	urldate = {2026-07-14},
	journal = {Monthly Notices of the Royal Astronomical Society},
	author = {Arur, K and Maccarone, T J},
	month = jun,
	year = {2022},
	pages = {1720--1732},
}

@article{Casella2005,
	title = {The {ABC} of {Low}-{Frequency} {Quasi}-periodic {Oscillations} in {Black} {Hole} {Candidates}: {Analogies} with {Z} {Sources}},
	volume = {629},
	issn = {0004-637X, 1538-4357},
	shorttitle = {The {ABC} of {Low}-{Frequency} {Quasi}-periodic {Oscillations} in {Black} {Hole} {Candidates}},
	url = {https://iopscience.iop.org/article/10.1086/431174},
	doi = {10.1086/431174},
	language = {en},
	number = {1},
	urldate = {2026-07-15},
	journal = {The Astrophysical Journal},
	author = {Casella, P. and Belloni, T. and Stella, L.},
	month = aug,
	year = {2005},
	pages = {403--407},
}

@article{Heil2015,
	title = {Inclination-dependent spectral and timing properties in transient black hole {X}-ray binaries},
	volume = {448},
	issn = {1365-2966, 0035-8711},
	url = {http://academic.oup.com/mnras/article/448/4/3348/976170/Inclinationdependent-spectral-and-timing},
	doi = {10.1093/mnras/stv240},
	language = {en},
	number = {4},
	urldate = {2026-07-14},
	journal = {Monthly Notices of the Royal Astronomical Society},
	author = {Heil, L. M. and Uttley, P. and Klein-Wolt, M.},
	month = apr,
	year = {2015},
	pages = {3348--3353},
}

@article{Ingram2009,
	title = {Low-frequency quasi-periodic oscillations spectra and {Lense}–{Thirring} precession},
	volume = {397},
	issn = {1745-3925, 1745-3933},
	url = {https://academic.oup.com/mnrasl/article/397/1/L101/1063411},
	doi = {10.1111/j.1745-3933.2009.00693.x},
	language = {en},
	number = {1},
	urldate = {2026-07-14},
	journal = {Monthly Notices of the Royal Astronomical Society: Letters},
	author = {Ingram, Adam and Done, Chris and Fragile, P. Chris},
	month = jul,
	year = {2009},
	pages = {L101--L105},
}

@article{Ingram2013,
	title = {An exact analytic treatment of propagating mass accretion rate fluctuations in {X}-ray binaries},
	volume = {434},
	issn = {0035-8711, 1365-2966},
	url = {http://academic.oup.com/mnras/article/434/2/1476/1070569/An-exact-analytic-treatment-of-propagating-mass},
	doi = {10.1093/mnras/stt1107},
	language = {en},
	number = {2},
	urldate = {2026-07-14},
	journal = {Monthly Notices of the Royal Astronomical Society},
	author = {Ingram, Adam and Klis, Michiel Van Der},
	month = sep,
	year = {2013},
	pages = {1476--1485},
}

@article{Kim1979,
	title = {Digital {Bispectral} {Analysis} and {Its} {Applications} to {Nonlinear} {Wave} {Interactions}},
	volume = {7},
	copyright = {https://ieeexplore.ieee.org/Xplorehelp/downloads/license-information/IEEE.html},
	issn = {0093-3813},
	url = {http://ieeexplore.ieee.org/document/4317207/},
	doi = {10.1109/TPS.1979.4317207},
	number = {2},
	urldate = {2026-07-14},
	journal = {IEEE Transactions on Plasma Science},
	author = {Kim, Young C. and Powers, Edward J.},
	year = {1979},
	pages = {120--131},
}

@article{Maccarone2002,
	title = {Higher order variability properties of accreting black holes},
	volume = {336},
	issn = {0035-8711, 1365-2966},
	url = {https://academic.oup.com/mnras/article/336/3/817/973442},
	doi = {10.1046/j.1365-8711.2002.05807.x},
	language = {en},
	number = {3},
	urldate = {2026-07-14},
	journal = {Monthly Notices of the Royal Astronomical Society},
	author = {Maccarone, T. J. and Coppi, P. S.},
	month = nov,
	year = {2002},
	pages = {817--825},
}

@article{Maccarone2011,
	title = {Coupling between {QPOs} and broad-band noise components in {GRS} 1915+105: {Coupled} variability in {GRS} 1915+105},
	volume = {413},
	issn = {00358711},
	shorttitle = {Coupling between {QPOs} and broad-band noise components in {GRS} 1915+105},
	url = {https://academic.oup.com/mnras/article-lookup/doi/10.1111/j.1365-2966.2011.18273.x},
	doi = {10.1111/j.1365-2966.2011.18273.x},
	language = {en},
	number = {3},
	urldate = {2026-07-14},
	journal = {Monthly Notices of the Royal Astronomical Society},
	author = {Maccarone, Thomas J. and Uttley, Philip and {van der Klis}, Michiel and Wijnands, Rudy A. D. and Coppi, Paolo S.},
	month = may,
	year = {2011},
	pages = {1819--1827},
}

@article{Motta2015,
	title = {Geometrical constraints on the origin of timing signals from black holes},
	volume = {447},
	issn = {1365-2966, 0035-8711},
	url = {http://academic.oup.com/mnras/article/447/2/2059/2593782/Geometrical-constraints-on-the-origin-of-timing},
	doi = {10.1093/mnras/stu2579},
	language = {en},
	number = {2},
	urldate = {2026-07-14},
	journal = {Monthly Notices of the Royal Astronomical Society},
	author = {Motta, S. E. and Casella, P. and Henze, M. and Muñoz-Darias, T. and Sanna, A. and Fender, R. and Belloni, T.},
	month = feb,
	year = {2015},
	pages = {2059--2072},
}

@article{Ponti2012,
	title = {Ubiquitous equatorial accretion disc winds in black hole soft states},
	volume = {422},
	issn = {1745-3933, 1745-3925},
	url = {https://academic.oup.com/mnrasl/article/422/1/L11/971072},
	doi = {10.1111/j.1745-3933.2012.01224.x},
	language = {en},
	number = {1},
	urldate = {2026-07-14},
	journal = {Monthly Notices of the Royal Astronomical Society: Letters},
	author = {Ponti, G. and Fender, R. P. and Begelman, M. C. and Dunn, R. J. H. and Neilsen, J. and Coriat, M.},
	month = may,
	year = {2012},
	pages = {L11--L15},
}

@article{TimmerKonig1995,
	title = {On generating power law noise},
	volume = {300},
	issn = {0004-6361},
	url = {https://ui.adsabs.harvard.edu/abs/1995A&A...300..707T},
	language = {en},
	urldate = {2026-07-15},
	journal = {Astronomy and Astrophysics},
	author = {Timmer, J. and König, M.},
	year = {1995},
	pages = {707--710},
}

@article{Uttley2005,
	title = {Non-linear {X}-ray variability in {X}-ray binaries and active galaxies},
	volume = {359},
	issn = {0035-8711, 1365-2966},
	url = {https://academic.oup.com/mnras/article-lookup/doi/10.1111/j.1365-2966.2005.08886.x},
	doi = {10.1111/j.1365-2966.2005.08886.x},
	language = {en},
	number = {1},
	urldate = {2026-07-14},
	journal = {Monthly Notices of the Royal Astronomical Society},
	author = {Uttley, P. and McHardy, I. M. and Vaughan, S.},
	month = may,
	year = {2005},
	pages = {345--362},
}

@ARTICLE{Wijnands1999,
       author = {{Wijnands}, Rudy and {Homan}, Jeroen and {van der Klis}, Michiel},
        title = "{The Complex Phase-Lag Behavior of the 3-12 HZ Quasi-Periodic Oscillations during the Very High State of XTE J1550-564}",
      journal = {\apjl},
         year = 1999,
        month = nov,
       volume = {526},
       number = {1},
        pages = {L33-L36},
          doi = {10.1086/312365},
archivePrefix = {arXiv},
       eprint = {astro-ph/9909515},
 primaryClass = {astro-ph},
       adsurl = {https://ui.adsabs.harvard.edu/abs/1999ApJ...526L..33W}
}

@article{Zhu2024b,
	title = {The {Bicoherence} {Analysis} of {Type}-{C} {Quasiperiodic} {Oscillations} in {Swift} {J1727}.8−1613},
	volume = {974},
	issn = {0004-637X, 1538-4357},
	url = {https://iopscience.iop.org/article/10.3847/1538-4357/ad7587},
	doi = {10.3847/1538-4357/ad7587},
	number = {2},
	urldate = {2026-07-14},
	journal = {The Astrophysical Journal},
	author = {Zhu, Haifan and Wang, Wei and Zhu, Ziyuan},
	month = oct,
	year = {2024},
	pages = {303},
}

@ARTICLE{Belloni2012,
       author = {{Belloni}, T.~M. and {Sanna}, A. and {M{\'e}ndez}, M.},
        title = "{High-frequency quasi-periodic oscillations in black hole binaries}",
      journal = {\mnras},
         year = 2012,
        month = nov,
       volume = {426},
       number = {3},
        pages = {1701-1709},
          doi = {10.1111/j.1365-2966.2012.21634.x},
archivePrefix = {arXiv},
       eprint = {1207.2311},
 primaryClass = {astro-ph.HE},
       adsurl = {https://ui.adsabs.harvard.edu/abs/2012MNRAS.426.1701B}
}

@ARTICLE{Remillard2002,
       author = {{Remillard}, Ronald A. and {Sobczak}, Gregory J. and {Muno}, Michael P. and {McClintock}, Jeffrey E.},
        title = "{Characterizing the Quasi-periodic Oscillation Behavior of the X-Ray Nova XTE J1550-564}",
      journal = {\apj},
         year = 2002,
        month = jan,
       volume = {564},
       number = {2},
        pages = {962-973},
          doi = {10.1086/324276},
archivePrefix = {arXiv},
       eprint = {astro-ph/0105508},
 primaryClass = {astro-ph},
       adsurl = {https://ui.adsabs.harvard.edu/abs/2002ApJ...564..962R}
}

@ARTICLE{Bellavita2022,
       author = {{Bellavita}, Candela and {Garc{\'\i}a}, Federico and {M{\'e}ndez}, Mariano and {Karpouzas}, Konstantinos},
        title = "{vKompth: a variable Comptonization model for low-frequency quasi-periodic oscillations in black hole X-ray binaries}",
      journal = {\mnras},
         year = 2022,
        month = sep,
       volume = {515},
       number = {2},
        pages = {2099-2109},
          doi = {10.1093/mnras/stac1922},
archivePrefix = {arXiv},
       eprint = {2206.13609},
 primaryClass = {astro-ph.HE},
       adsurl = {https://ui.adsabs.harvard.edu/abs/2022MNRAS.515.2099B}
}

@ARTICLE{Stella1998,
       author = {{Stella}, Luigi and {Vietri}, Mario},
        title = "{Lense-Thirring Precession and Quasi-periodic Oscillations in Low-Mass X-Ray Binaries}",
      journal = {\apjl},
         year = 1998,
        month = jan,
       volume = {492},
       number = {1},
        pages = {L59-L62},
          doi = {10.1086/311075},
archivePrefix = {arXiv},
       eprint = {astro-ph/9709085},
 primaryClass = {astro-ph},
       adsurl = {https://ui.adsabs.harvard.edu/abs/1998ApJ...492L..59S}
}

@ARTICLE{Marcel2026,
       author = {{Marcel}, G. and {Turner}, S.~G.~D. and {Ricketts}, B.~J. and {L{\'o}pez-Barquero}, V. and {Buisson}, D.~J.~K. and {Vincentelli}, F. and {Middleton}, M. and {Reynolds}, C.~S. and {Avara}, M.~J.},
        title = "{Disk warping and black hole X-ray binaries: I. Tentative unification of low-frequency quasi-periodic oscillations}",
      journal = {\aap},
         year = 2026,
        month = jul,
       volume = {710},
          eid = {A387},
        pages = {A387},
          doi = {10.1051/0004-6361/202558103},
archivePrefix = {arXiv},
       eprint = {2511.10474},
 primaryClass = {astro-ph.HE},
       adsurl = {https://ui.adsabs.harvard.edu/abs/2026A&A...710A.387M}
}

@phdthesis{Fackrell1997,
  author  = {Fackrell, Justin W. A.},
  title   = {Bispectral Analysis of Speech Signals},
  school  = {University of Edinburgh},
  address = {Edinburgh, UK},
  year    = {1996},
  month   = sep,
  type    = {{Ph.D.} thesis},
  note    = {School of Engineering and Electronics. Supervisor: S. McLaughlin},
  url     = {http://hdl.handle.net/1842/1384}
}

@book{Proakis1992,
	address = {New York, NY},
	title = {Advanced digital signal processing},
	isbn = {978-0-02-396841-9},
	language = {eng},
	publisher = {Macmillan},
	author = {Proakis, John G. and Rader, Charles M. and Ling, Fuyun and Nikias, Chrysostomos L.},
	year = {1992},
}

@article{Motta2011,
	title = {Low frequency oscillations in black holes: a spectral-timing approach to the case of {GX} 339-4},
	volume = {418},
	issn = {00358711},
	url = {http://arxiv.org/abs/1108.0540},
	doi = {10.1111/j.1365-2966.2011.19566.x},
	number = {4},
	journal = {Monthly Notices of the Royal Astronomical Society},
	author = {Motta, S. and Muñoz-Darias, T. and Casella, P. and Belloni, T. and Homan, J.},
	month = dec,
	year = {2011},
	note = {arXiv:1108.0540 [astro-ph.HE]},
	pages = {2292--2307},
}

@article{Ingram2011,
	title = {A physical model for the continuum variability and {QPO} in accreting black holes},
	volume = {415},
	issn = {00358711},
	url = {http://arxiv.org/abs/1101.2336},
	doi = {10.1111/j.1365-2966.2011.18860.x},
	number = {3},
	journal = {Monthly Notices of the Royal Astronomical Society},
	author = {Ingram, Adam and Done, Chris},
	month = aug,
	year = {2011},
	note = {arXiv:1101.2336 [astro-ph.SR]},
	pages = {2323--2335},
}

@article{Ingram2019,
	title = {A review of quasi-periodic oscillations from black hole {X}-ray binaries: observation and theory},
	volume = {85},
	issn = {13876473},
	url = {http://arxiv.org/abs/2001.08758},
	doi = {10.1016/j.newar.2020.101524},
	journal = {New Astronomy Reviews},
	author = {Ingram, Adam and Motta, Sara},
	month = sep,
	year = {2019},
	note = {arXiv:2001.08758 [astro-ph.HE]},
	pages = {101524},
}

@ARTICLE{Remillard2006,
       author = {{Remillard}, Ronald A. and {McClintock}, Jeffrey E.},
        title = "{X-Ray Properties of Black-Hole Binaries}",
      journal = {\araa},
         year = 2006,
        month = sep,
       volume = {44},
       number = {1},
        pages = {49-92},
          doi = {10.1146/annurev.astro.44.051905.092532},
archivePrefix = {arXiv},
       eprint = {astro-ph/0606352},
 primaryClass = {astro-ph},
       adsurl = {https://ui.adsabs.harvard.edu/abs/2006ARA&A..44...49R}
}

@ARTICLE{Wirnitzer1985,
       author = {{Wirnitzer}, B.},
        title = "{Bispectral analysis at low light levels and astronomical speckle masking.}",
      journal = {Journal of the Optical Society of America A},
         year = 1985,
        month = jan,
       volume = {2},
        pages = {14-21},
          doi = {10.1364/JOSAA.2.000014},
       adsurl = {https://ui.adsabs.harvard.edu/abs/1985JOSAA...2...14W}
}

@unpublished{Jin2026inprep,
    author = {{Jin}, Pei and {M{\'e}ndez}, Mariano and {Garc{\'i}a}, Federico and {Altamirano}, Diego and {Zhu}, Haifan and {Zhang}, Liang and {Zhang}, Guobao},
    title = "{A possible connection between the ridge and hypotenuse in the bispectrum of black hole X-ray binaries through quasi-periodic modulation}",
    note = {Submitted to Monthly Notices of the Royal Astronomical Society},
    year = {2026, submitted concurrently},
}

@ARTICLE{Parra2025,
       author = {{Parra}, M. and {Bianchi}, S. and {Petrucci}, P.-O. and {Bouchet}, T. and {Shidatsu}, M. and {Capitanio}, F. and {Dov{\v{c}}iak}, M. and {Russell}, T.~D. and {Gianolli}, V.~E. and {Carotenuto}, F.},
        title = "{20 years of disk winds in 4U 1630{\ensuremath{-}}47: I. Long-term behavior and influence of hard X-rays}",
      journal = {\aap},
         year = 2025,
        month = sep,
       volume = {701},
          eid = {A292},
        pages = {A292},
          doi = {10.1051/0004-6361/202554976},
archivePrefix = {arXiv},
       eprint = {2504.00991},
 primaryClass = {astro-ph.HE},
       adsurl = {https://ui.adsabs.harvard.edu/abs/2025A&A...701A.292P}
}

@ARTICLE{Shidatsu2026,
       author = {{Shidatsu}, Megumi and {Parra}, Maxime},
        title = "{X-ray Spectroscopy of Disk Winds in Black Hole X-ray Binaries}",
      journal = {arXiv e-prints},
         year = 2026,
        month = jan,
          eid = {arXiv:2601.22836},
        pages = {arXiv:2601.22836},
          doi = {10.48550/arXiv.2601.22836},
archivePrefix = {arXiv},
       eprint = {2601.22836},
 primaryClass = {astro-ph.HE},
       adsurl = {https://ui.adsabs.harvard.edu/abs/2026arXiv260122836S}
}

@ARTICLE{Sanchez2024,
       author = {{Mata S{\'a}nchez}, D. and {Mu{\~n}oz-Darias}, T. and {Armas Padilla}, M. and {Casares}, J. and {Torres}, M.~A.~P.},
        title = "{Evidence for inflows and outflows in the nearby black hole transient Swift J1727.8‒162}",
      journal = {\aap},
         year = 2024,
        month = feb,
       volume = {682},
          eid = {L1},
        pages = {L1},
          doi = {10.1051/0004-6361/202348754},
archivePrefix = {arXiv},
       eprint = {2401.04107},
 primaryClass = {astro-ph.HE},
       adsurl = {https://ui.adsabs.harvard.edu/abs/2024A&A...682L...1M}
}

@article{2007CSE.....9c..21P,
  author        = {{Perez}, Fernando and {Granger}, Brian E.},
  title         = "{IPython: A System for Interactive Scientific Computing}",
  journal       = {Computing in Science and Engineering},
  year          = "2007",
  month         = "Jan",
  volume        = {9},
  number        = {3},
  pages         = {21--29},
  doi           = {10.1109/MCSE.2007.53},
  adsurl        = {https://ui.adsabs.harvard.edu/abs/2007CSE.....9c..21P}
}

@conference{kluyver2016jupyter,
  title         = {Jupyter Notebooks -- a publishing format for reproducible computational workflows},
  author        = {Thomas Kluyver and Benjamin Ragan-Kelley and Fernando P{\'e}rez and Brian Granger and Matthias Bussonnier and Jonathan Frederic and Kyle Kelley and Jessica Hamrick and Jason Grout and Sylvain Corlay and Paul Ivanov and Dami{\'a}n Avila and Safia Abdalla and Carol Willing},
  booktitle     = {Positioning and Power in Academic Publishing: Players, Agents and Agendas},
  editor        = {F. Loizides and B. Schmidt},
  organization  = {IOS Press},
  pages         = {87--90},
  year          = {2016}
}

@article{Hunter:2007,
  author        = {Hunter, J. D.},
  title         = {Matplotlib: A 2D graphics environment},
  journal       = {Computing in Science \& Engineering},
  volume        = {9},
  number        = {3},
  pages         = {90--95},
  publisher     = {IEEE COMPUTER SOC},
  doi           = {10.1109/MCSE.2007.55},
  year          = 2007
}

@article{numpy,
  title         = {Array programming with {NumPy}},
  author        = {Charles R. Harris and K. Jarrod Millman and St{\'{e}}fan J. van der Walt and Ralf Gommers and Pauli Virtanen and David Cournapeau and Eric Wieser and Julian Taylor and Sebastian Berg and Nathaniel J. Smith and Robert Kern and Matti Picus and Stephan Hoyer and Marten H. van Kerkwijk and Matthew Brett and Allan Haldane and Jaime Fern{\'{a}}ndez del R{\'{i}}o and Mark Wiebe and Pearu Peterson and Pierre G{\'{e}}rard-Marchant and Kevin Sheppard and Tyler Reddy and Warren Weckesser and Hameer Abbasi and Christoph Gohlke and Travis E. Oliphant},
  year          = {2020},
  month         = sep,
  journal       = {Nature},
  volume        = {585},
  number        = {7825},
  pages         = {357--362},
  doi           = {10.1038/s41586-020-2649-2},
  publisher     = {Springer Science and Business Media {LLC}},
  url           = {https://doi.org/10.1038/s41586-020-2649-2}
}

@book{python,
  author        = {Van Rossum, Guido and Drake, Fred L.},
  title         = {Python 3 Reference Manual},
  year          = {2009},
  isbn          = {1441412697},
  publisher     = {CreateSpace},
  address       = {Scotts Valley, CA}
}
\bibliographystyle{aasjournalv7}

\end{document}